\documentclass[a4paper]{jpconf}
\usepackage{graphicx}
\begin{document}
\title{Compact Yb$^+$ optical atomic clock project: design principle  and current status}

\author{Cl\'ement Lacro\^ute, Ma\"el Souidi, Pierre-Yves Bourgeois, Jacques Millo, Khaldoun Saleh, Emmanuel Bigler, Rodolphe Boudot, Vincent Giordano and Yann Kersal\'e}

\address{FEMTO-ST, TF Dpt., UMR 6174 CNRS-ENSMM-UFC-UTBM, 26 ch de l'Epitaphe, 25030 Besan\c{c}on, France}

\ead{clement.lacroute@femto-st.fr}

\begin{abstract}
We present the design of a compact optical clock based on the $^2S_{1/2} \rightarrow ^2D_{3/2}$ 435.5~nm transition in $^{171}$Yb$^+$. The ion trap will be based on a micro-fabricated circuit, with surface electrodes generating a trapping potential to localize a single Yb ion a few hundred $\mu$m from the electrodes. We present our trap design as well as simulations of the resulting trapping pseudo-potential.  We also present a compact, multi-channel wavelength meter that will permit the frequency stabilization of the cooling, repumping and clear-out lasers at 369.5~nm, 935.2~nm and 638.6~nm needed to cool the ion. We use this wavelength meter to characterize and stabilize the frequency of extended cavity diode lasers at 369.5~nm and 638.6~nm.
\end{abstract}

\section{Introduction}
Optical atomic clocks have now surpassed atomic fountain clocks by two orders of magnitude both in terms of fractional frequency stability and systematic uncertainty \cite{Bloom2014, Ushijima2015}. They have proven to be useful for frequency metrology, for fundamental physics tests \cite{Huntemann2014, Godun2014}, and their resolution make them competitive sensors for relativistic geodesy \cite{Hinkley2013, Margolis2013}. While theoretical proposals point out new ways to further improve their performances \cite{Yudin2010, Derevianko2012, Zanon-Willette2015}, and a new defintion of the SI second by an optical transition is being discussed \cite{Riehle2015}, interest starts to arise for field measurements and transportable atomic clocks \cite{Margolis2014}.

Several proposals for fundamental physics tests using space optical clocks have been written, and prototypes of such clocks are being developed \cite{Schiller2008, Bongs2015}. Time distribution through GNSS would benefit both from optical clocks in the satellites and in the ground stations \cite{Moudrak2008}. Additionally, one of the most immediate benefits of a transportable optical clock would be centimetric resolution for relativistic geodesy surveys \cite{Delva2013}.

In this perspective, we are currently designing a compact, single-ion atomic clock based on the 435.5 nm transition in trapped $^{171}$Yb$^+$, with a target frequency stability one order of magnitude below hydrogen masers in a comparable volume, \emph{e.g.} $10^{-14}\tau^{-1/2}$ in about 500 L. Yb$^+$ can be manipulated with diode lasers, and the 435.5 nm transition can easily be adressed by second-harmonic generation, which can be integrated in a compact optical setup. The clock will be based on a surface electrodes (SE) trap \cite{Chiaverini2005}. SE traps are now widely used in quantum information processing (QIP) experiments. They allow for a compact physics package, benefit from high-end microfabrication techniques, and allow for the integration of optical elements \cite{Eltony2015}. They are known to be hindered by the so-called ``anomalous heating" \cite{Turchette2000}, due to the proximity of the trapping electrodes that can radiate patch potentials \cite{Dubessy2009}, but a way to circumvent this phenomenon using ion-bombardment of the electrodes was recently found \cite{Hite2012, McKay2014}, making such traps compatible with a metrological setup.

In this article, we present the general design of the clock, with an emphasis on the SE trap and the laser frequency stabilization of extended cavity diode lasers using a multi-channel, fibered wavelength meter.

\section{Surface electrode trap ion clock: general principle}

\subsection{Clock design}
\begin{figure}[h!]
   \begin{minipage}[c]{.46\linewidth}
			\centering
      \includegraphics[width=1.2\columnwidth]{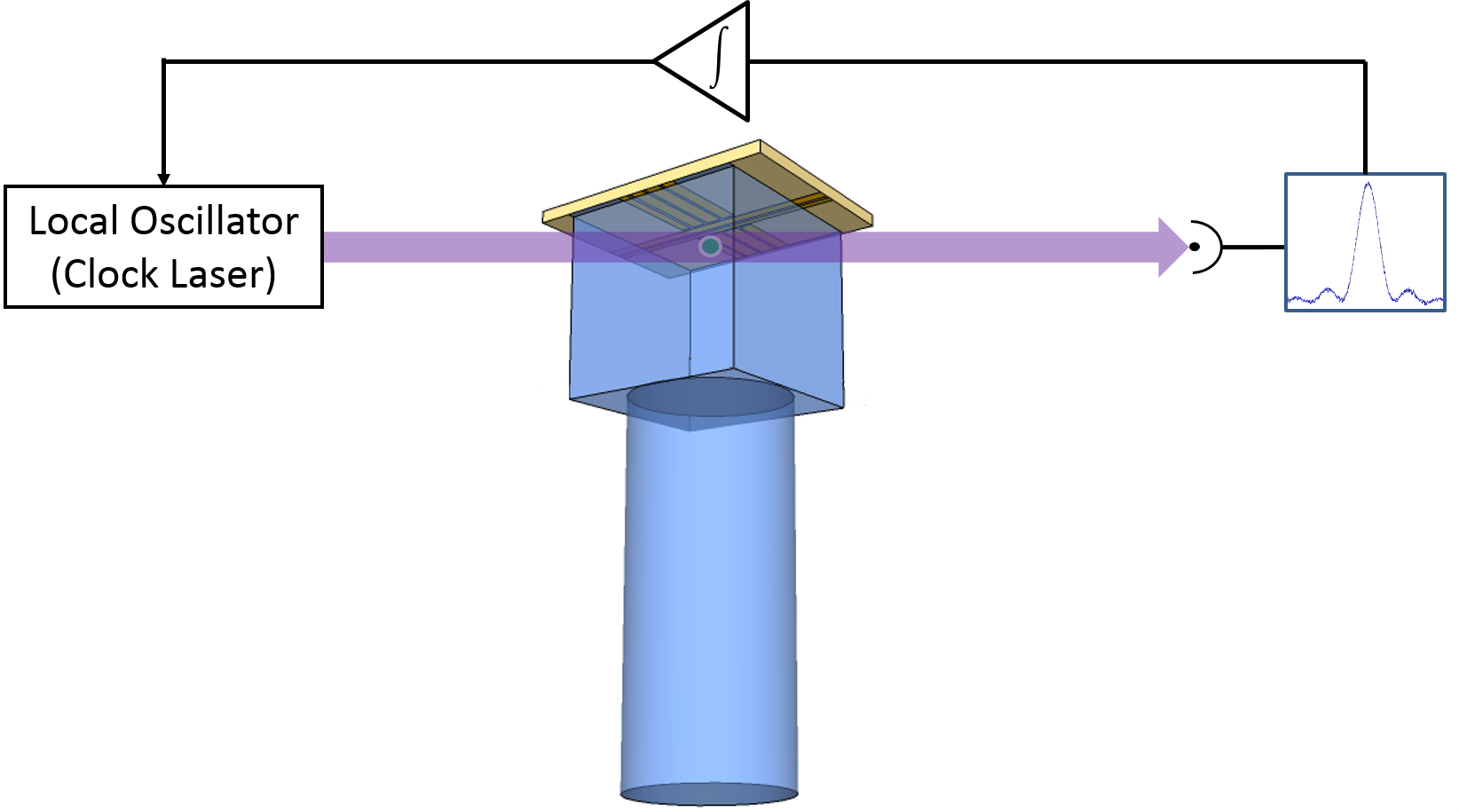}
			\caption{Schematic view of the experiment. A surface electrode trap confines an Yb$^+$ ion 600 $\mu$m below the chip surface. The chip closes a cubic (1 inch)$^3$ glass cell evacuated to ultra-high vacuum. The clock laser probes the clock transition and is locked to the atomic signal.}
			\label{fig:ppe}
   \end{minipage} \hfill
   \begin{minipage}[c]{.46\linewidth}
			\centering
      \includegraphics[width=0.6\columnwidth]{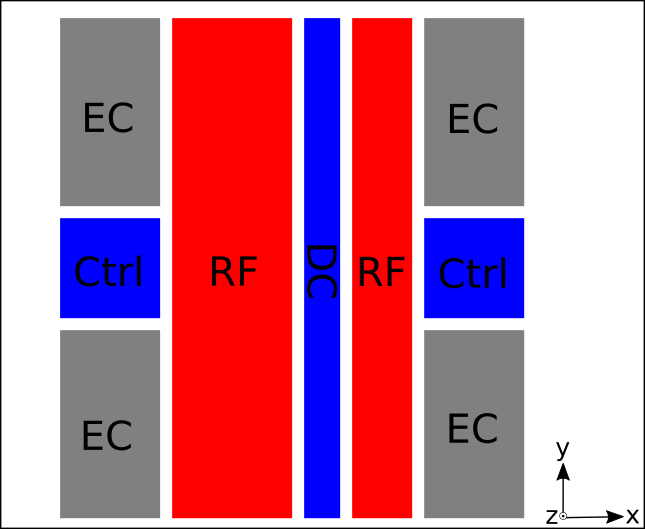}
			\caption{Schematic of the five-wire trap design. RF: radio-frequency electrodes. Ctrl: DC control electrodes. DC: central control electrodes. EC: DC endcap electrodes.}%
			\label{fig:trap}
   \end{minipage}
\end{figure}

We aim to build a compact atomic clock with a total volume of order 500 L, enabled by the use of a surface electrode ion trap combined with a compact vacuum chamber analogous to those used in ``atom chips'' experiments \cite{Du2004}. The trap will close a cubic glass cell connected to a hybrid ion/getter pump, see a preliminary schematic illustration figure \ref{fig:ppe}. Such vacuum chamber designs have been used in compact ($<$~1\ m$^3$) Bose-Einstein condensation setups \cite{Farkas2010}. The pressure obtained in such setups is in the low 10$^{-10}$ mbars range, in principle compatible with Ytterbium ion trapping and cooling. To our knowledge, such ``atom chips'' cells have never been tested in ion trapping setups, and we will adapt and test such a design with Yb$^+$ ions.

The cooling and ionization lasers will be stabilized using a compact, multi-channel wavelength meter (see section \ref{sec:wm2}). The clock laser will be generated by frequency doubling of an extended cavity diode laser (ECDL) at 871 nm. Second harmonic generation will be obtained with a fibered non-linear optical waveguide, avoiding the use of a resonant cavity. It will be pre-stabilized near the electric quadrupole transition at 435.5 nm using an ultra-stable optical cavity; we are developping an ultra-compact ULE Fabry-Perot cavity at telecom wavelength, with a projected total volume of about 54 L for the ultra-stable laser system. We will develop a similar cavity at 871 nm, in order to reduce the ultra-stable clock laser volume as much as possible.

\subsection{Trap design}

In most metrological experiments with ions, 3D designs are used for the trapping electrodes. Similar trapping potentials can be obtained using surface electrodes traps. The main difference comes from the plane shaped electrodes which are all in the same plane and generate the confinement field above the surface of the trap. The pseudo-potential generated by surface electrodes can be computed analytically \cite{House2008, Wesenberg2008, Chiaverini2005} and the field can be approximated as a quadrupolar one just near the center of the trap. Due to low electrodes radio-frequency (RF) breakdown voltages, typically close to 300 V, SE traps are usually weaker than traditional Paul traps. 

Our system is a five wire RF SE trap illustrated in figure \ref{fig:trap}. It consists of two RF electrodes separated by a DC control electrode and surrounded by two segmented DC electrodes (the ``end caps"). Our two RF electrodes are asymmetric, one of the electrodes being wider than the other (by a factor of two) in order to rotate the principal axes of the trap compared to the surface. The resulting tilt is about 8$^\circ$ with regard to the chip frame. This will allow us to laser cool all the motional axes of our ion even when the laser beam is parallel to the electrodes.

We aim to maximise the ion trapping distance from the electrodes, $d$, as the so-called ``anomalous heating" which affects SE traps scales with $d^{-4}$ \cite{Deslauriers2006}. In order to obtain at least 500 $\mu$m between the ion and the surface of the trap, the two RF electrodes are respectively 600 $\mu$m and 1200 $\mu$m wide (the distance between the ion and the trap surface is related to the electrodes width and spacing, see for instance \cite{Chiaverini2005}). The central DC control electrode is 600 $\mu$m wide, and the length of these three electrodes is about 5 mm. These dimensions locate the center of the saddle point at 580 $\mu$m above the surface of the trap and rotates the principal radial axes of motion. At this distance, we can expect a normalized electric field noise power spectral density $\omega S_E(\omega)\approx 4{\times}10^{-6}\ \mathrm{V^2/m^2}$ \cite{Hite2012}. The resulting heating rate for a trapping frequency of 100~kHz is of order 2200 phonons/s $\approx 20$ mK/s. Even though this is rather high, the corresponding temperature rise during a probe pulse of 50 ms is about 1 mK, corresponding to an additional second-order Doppler shift of about $3 {\times} 10^{-19}$. Moreover, this heating rate can be reduced by a factor of up to 100 using surface cleaning by an Ar beam \cite{Hite2012}, which would make this effect completely negligible.

\begin{figure}[h!]%
\centering
\includegraphics[width=0.4\columnwidth]{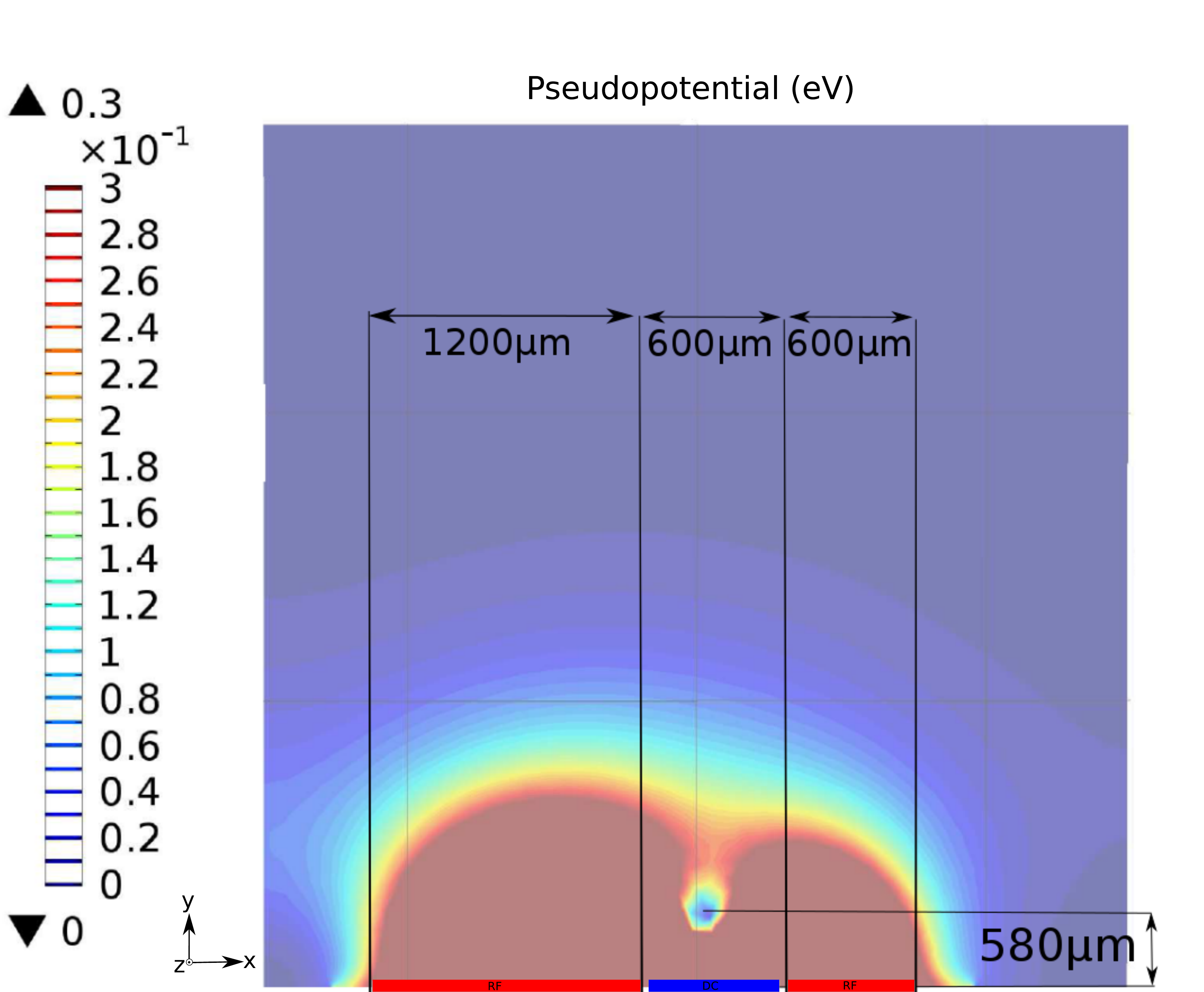}%
\includegraphics[width=0.38\columnwidth]{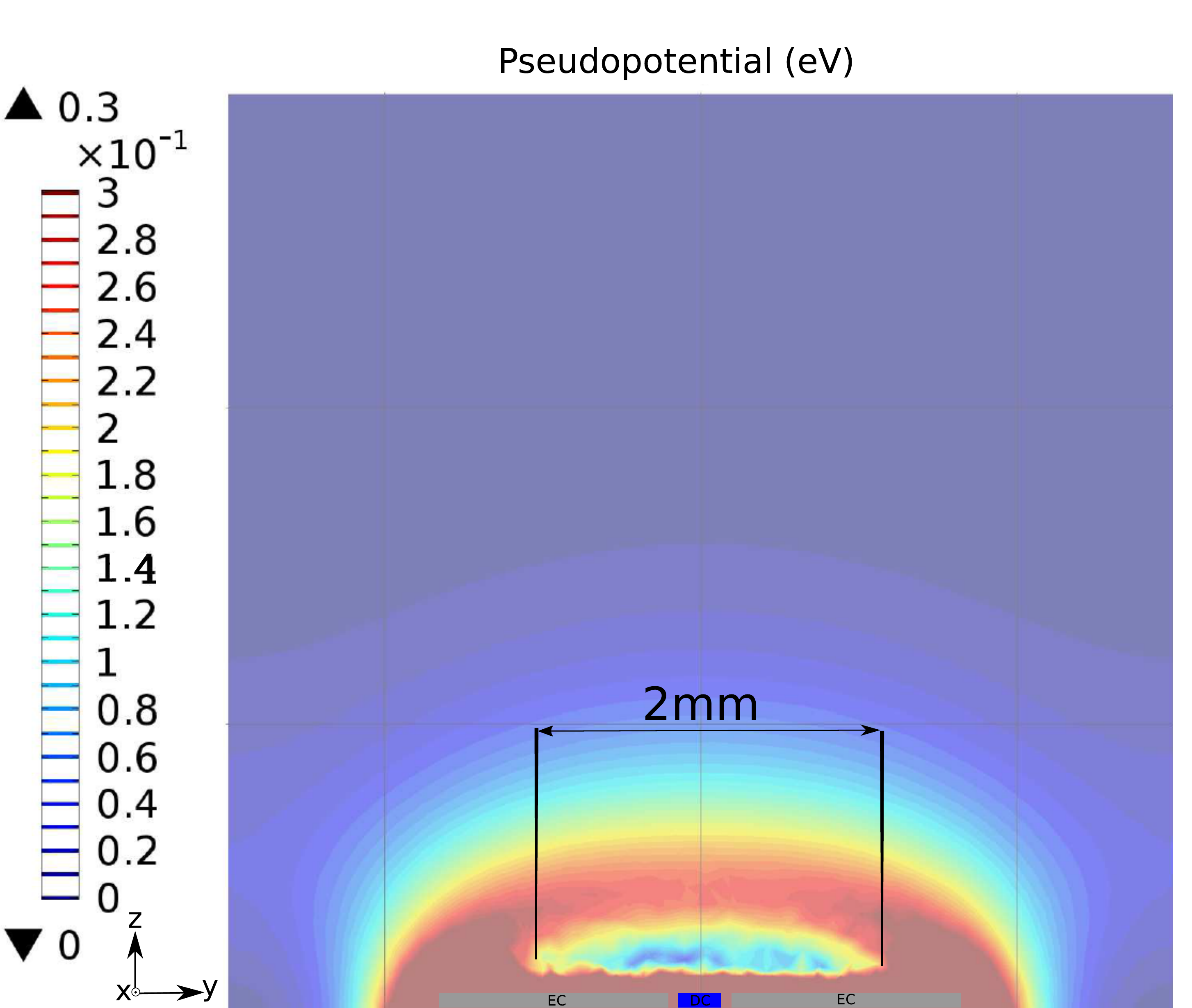}%
\caption{FEM simulations of the pseudopotential generated by our set of electrodes. \emph{Left}: the ion will be localized at 580 $\mu$m above the trap surface. The axis of minimal energy is tilted compared to the vertical axis. This effect is due to the asymmetry of the RF electrodes widths. \emph{Right}: pseudopotential, showing the trap in the longitudinal direction.}%
\label{fig:potential}%
\end{figure}

We also want a trap depth higher than 200 meV and a radial stability parameter $q_r < 0.3$. Following simulations of our trap geometry based on \cite{House2008}, we decided to apply an RF voltage of 250 V at 5.5 MHz. That leads us to a theoretical depth of 222 meV and a radial stability parameter $q_r=0.27$. By computing the curvature of the pseudo-potential in the center of the trap we find an oscillating motion frequency of 500 kHz in the radial plane and 97 kHz along the longitudinal axis. Figure \ref{fig:potential} shows finite-element method (FEM) simulations of the pseudo-potential.

\section{Characterization and wavelength stabilization of extended cavity diode lasers using a Fizeau wavelength meter} \label{sec:wm2}

Ultra-violet (UV) laser diodes are still difficult to realize, and only few wavelengths are available below 400 nm. The 369.52 nm cooling transition in Yb$^+$ is therefore often obtained by frequency doubling a 740 nm extended cavity diode laser, which can be stabilized to a Fabry Perot transfer cavity \cite{McLoughlin2011}, an iodine cell, or a wavelength meter \cite{McLoughlin2011,Pyka2013}. More recently, diodes emitting directly around 370 nm near room-temperature have become available, and such diodes can either be stabilized to Yb$^+$ using a discharge lamp \cite{Lee2014}, or to a wavelength meter, as demonstrated here.

The transitions involved in $^{171}$Yb$^+$ cooling have linewidth $\gamma$ of order 10 MHz. The integrated laser frequency fluctuations must be well below $\gamma$ on the laser-cooling timescale, which is about a few ten ms. This translates in a fractional frequency stability well below $10^{-9}$ at short term for the wavelengths which are considered here (369.5 nm and 638 nm). For a clock operated on a daily basis, the lasers should also have negligible drift rates (below 1 MHz/day) to stay on resonance on the longer term.

\subsection{Wavelength meter performance}

The wavelength meter we use for laser frequency stabilization is a Standard WS7 wavelength meter from HighFinesse, with a wavelength measurement range from 350 nm to 1120 nm. An optional multichannel fiber-switch unit can be used to  measure up to eight lasers and an optional proportional integral (PI) regulator can be used to perform a frequency-lock of all these lasers. We plan to use the 4-channel version of the switch to simultaneously measure and stabilize the cooling, clear-out and ionization lasers at 369.5 nm, 638 nm and 398 nm. For the measurements presented here, only one channel is used.

We have characterized the wavelength meter in terms of fractional frequency stability using the experimental setup described in \cite{Saleh2015}. We calculated the Allan deviation of the measured wavelength (771 nm) when using an ultra-stable, frequency-doubled, 1542 nm laser as an input. The ultra-stable laser exhibits a fractional frequency instability of about $2{\times}10^{-15}$ at 1s and a drift rate of 6 kHz/day \cite{Didier2015}, far below the wavelength meter instability. The resulting Allan deviation is therefore the wavelength meter fractional frequency stability. We obtain a fractional frequency instability of $2{\times}10^{-10} \ \tau^{-1/2}$ from 0.15 s to 10 s and a long-term frequency drift of $3{\times}10^{-13} \ \tau$ (about 10 MHz/day) at 771 nm. The resulting relative frequency instability is shown in figures \ref{fig:stab} and \ref{fig:stab2}, scaled to 369.52 nm and 638.61 nm, assuming that the resolution is the same at all wavelengths  (dashed black line).

 

\subsection{Frequency control with the wavelengthmeter}

\begin{figure}
   \begin{minipage}[c]{.46\linewidth}
			\centering
      \includegraphics[width=1\columnwidth]{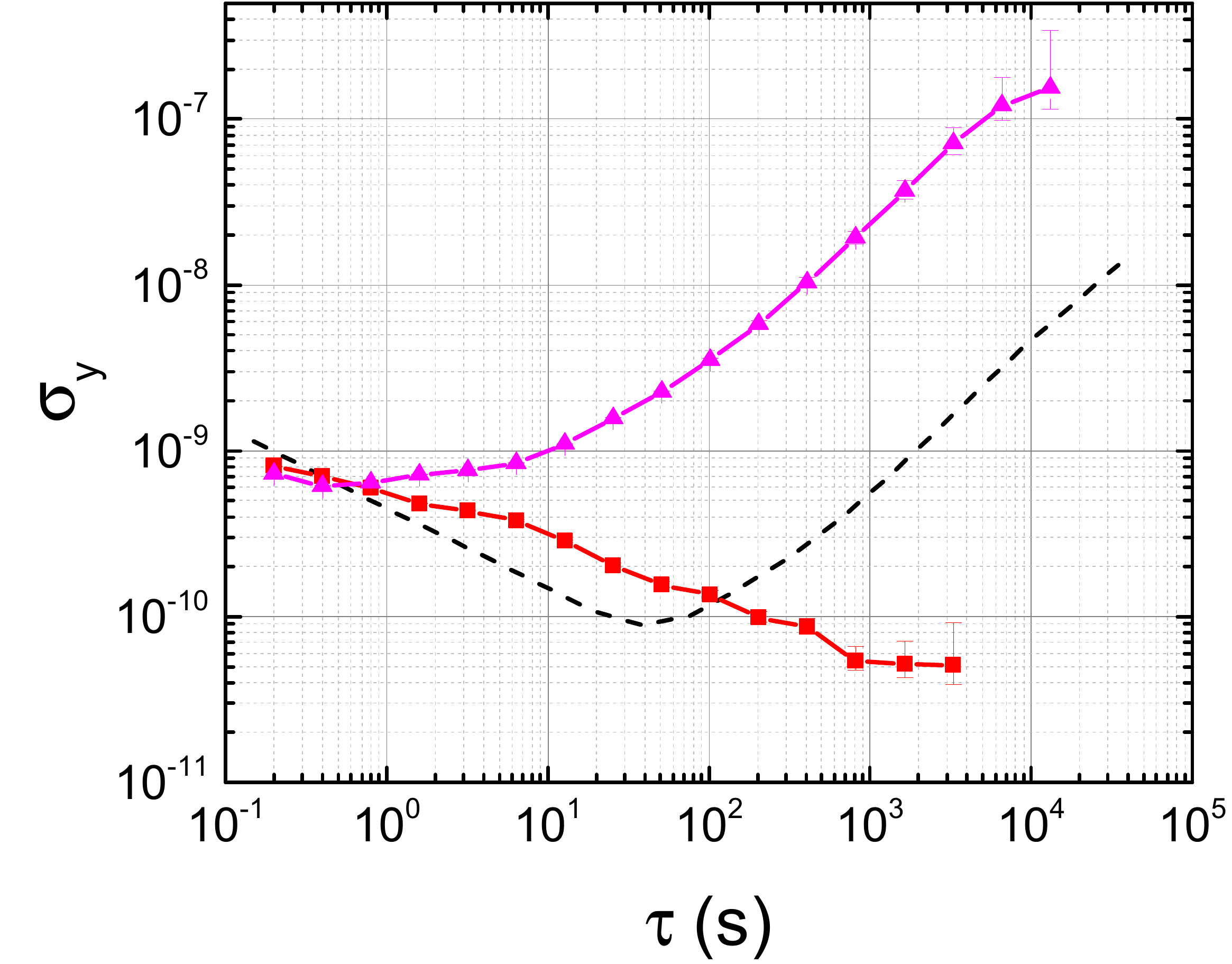}
			\caption{Characterization and frequency stabilization of the 369.52 nm cooling laser by the wavelength meter (WM). \textit{Dashed line:} stability of the wavelength meter, scaled to 369.52 nm. \textit{Pink triangles:} free-running Extended Cavity Diode Laser (ECDL) at 369.52 nm. \textit{Red squares:} ECDL at 369.52 nm locked to the WM using the proprietary software, in-loop signal.}
			\label{fig:stab}
   \end{minipage} \hfill
   \begin{minipage}[c]{.46\linewidth}
			\centering
      \includegraphics[width=1\columnwidth]{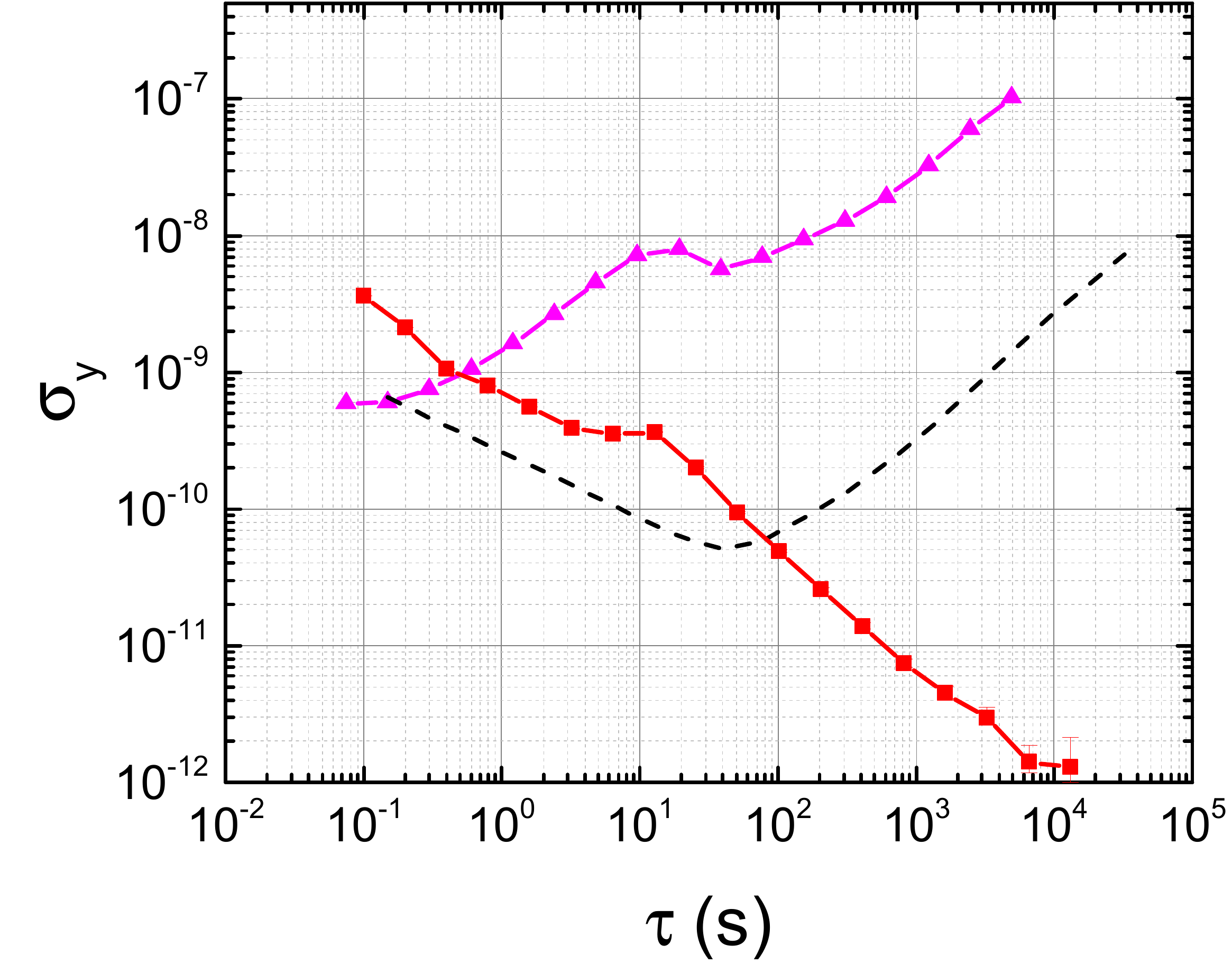}
			\caption{Characterization and frequency stabilization of the 638.61 nm cooling laser by the wavelength meter (WM). \textit{Dashed line:} stability of the wavelength meter, scaled to 638.61 nm. \textit{Pink triangles:} free-running Extended Cavity Diode Laser at 638.1 nm. \textit{Red squares:} ECDL at 638.61 nm locked to the WM using our own digital PI loop, in-loop signal, outliers removed.}%
			\label{fig:stab2}
   \end{minipage}
\end{figure}

We have first calculated the Allan deviation of a free-running ECDL wavelength at 369.5 nm, as shown in figure \ref{fig:stab} (pink triangles). The measured fractional frequency stability is below $10^{-9}$ for $\tau$ between 0.1 and 10 s, which is typical for ECDLs \cite{Rovera1994}. The red squares show the locked laser in-loop fractional frequency stability as measured by the wavelength meter when using the wavelength meter proprietary software. The dashed line indicates the best performance achievable by the locked laser, assuming that the wavelength meter resolution is the same at 771 nm and 369.52 nm. It is also important to note that the ECDL and the wavelength meter were not characterized at the same time, nor in the same room, which most probably affects the results. Assessing the ECDL locked performance would necessitate an independent reading of its wavelength, which we cannot perform at the moment.

Nevertheless, it is fair to state that the best possible performance achievable by the locked laser is about $5{\times}10^{-10}$ at 1 s, with a 30-fold improvement over its frequency drift.

The ion clock operation will require a complex control system, with inter-dependent parameters that will need to be optimized. We will use homebuilt digital electronics whenever possible, both for compacity and for the high degree of control allowed by multiple input-multiple output (MIMO) control systems. In that spirit, we have developped our own frequency control loop using the wavelength meter. The wavelength read by the wavelength meter is transferred via ethernet to a digital PI controller including a digital to analog converter and a FPGA. The correction signal acts on the piezoelectric transducer of the ECDL. Figure \ref{fig:stab2} shows preliminary results obtained when locking the clear-out laser at 638.61 nm. The free-running laser shows excess noise around $\tau=10$ s, probably due to temperature fluctuations. Even though this noise is not fully rejected, the resulting in-loop stability is below $10^{-9}$ for $\tau >$ 0.5 s. We suspect that the short-term excess frequency noise stems from communication timing jitter, as the data transits from the wavemeter server computer to the PI board. Once this problem is solved and with fully optimized gain parameters, stability below $3{\times}10^{-10}$  from 1 to 1000 s should be achievable.

It is worth noting that the drift rate of about 10 MHz/day only allows autonomous operation on the day scale. This drift should therefore be compensated either numerically, or by using a slower servo loop based on the trapped ions fluorescence signal.

\section{Conclusion}
In conclusion, we have presented the status of our compact Yb$^+$ optical clock project. A testbed SE trap has been designed. The ion will be trapped at a distance greater than 500 $\mu$m in order to reduce the effect of the so-called anomalous heating. Transverse trapping greater than 500~kHz will ensure operation in the Lamb-Dicke regime.

The compact clock will of course incorporate a compact optical bench. The cooling lasers frequency stabilization will be performed using a multichannel, fibered wavelength meter. We have measured the wavelength meter frequency stability, and used it to characterize and lock the cooling and clear-out ECDLs at 369.52~nm and 638.61~nm. The performances are compatible with ion cooling and detection requirements.

This work is funded by the Agence Nationale de la Recherche (ANR) (ANR-14-CE26-0031-01-MITICC); ANR Programme d'Investissement d'Avenir (PIA) (First-TF network, Oscillator-IMP project); and the R\'egion Franche Comt\'e.

\section*{References}

\bibliographystyle{iopart-num}
\bibliography{fsm2015}

\providecommand{\newblock}{}
\begin{thebibliography}{10}
\expandafter\ifx\csname url\endcsname\relax
  \def\url#1{{\tt #1}}\fi
\expandafter\ifx\csname urlprefix\endcsname\relax\def\urlprefix{URL }\fi
\providecommand{\eprint}[2][]{\url{#2}}

\bibitem{Bloom2014}
Bloom B~J, Nicholson T~L, Williams J~R, Campbell S~L, Bishof M, Zhang X, Zhang
  W, Bromley S~L and Ye J 2014 {\em Nature\/} {\bf 506} 71--75

\bibitem{Ushijima2015}
Ushijima I, Takamoto M, Das M, Ohkubo T and Katori H 2015 {\em Nat. Phot.\/}
  {\bf 9} 185--189

\bibitem{Huntemann2014}
Huntemann N, Lipphardt B, Tamm C, Gerginov V, Weyers S and Peik E 2014 {\em
  Phys. Rev. Lett.\/} {\bf 113} 210802

\bibitem{Godun2014}
Godun R, Nisbet-Jones P, Jones J, King S, Johnson L, Margolis H, Szymaniec K,
  Lea S, Bongs K and Gill P 2014 {\em Phys. Rev. Lett.\/} {\bf 113} 210801

\bibitem{Hinkley2013}
Hinkley N, Sherman J~A, Phillips N~B, Schioppo M, Lemke N~D, Beloy K, Pizzocaro
  M, Oates C~W and Ludlow A~D 2013 {\em Science\/} {\bf 341} 1215--1218

\bibitem{Margolis2013}
Margolis H~S 2013 {\em Science\/} {\bf 341} 1185--1186

\bibitem{Yudin2010}
Yudin V~I, Taichenachev A~V, Oates C~W, Barber Z~W, Lemke N~D, Ludlow A~D,
  Sterr U, Lisdat C and Riehle F 2010 {\em Phys. Rev. A\/} {\bf 82} 011804

\bibitem{Derevianko2012}
Derevianko A, Dzuba V~A and Flambaum V~V 2012 {\em Phys. Rev. Lett.\/} {\bf
  109} 180801

\bibitem{Zanon-Willette2015}
Zanon-Willette T, Yudin V~I and Taichenachev A~V 2015 {\em Phys. Rev. A\/} {\bf
  92} 023416

\bibitem{Riehle2015}
Riehle F 2015 {\em Comptes Rendus Physique\/} {\bf 16} 506--515

\bibitem{Margolis2014}
Margolis H 2014 {\em Nat. Phys.\/} {\bf 10} 82--83

\bibitem{Schiller2008}
{Schiller S \emph{et al}} 2008 {\em Exp. Astro.\/} {\bf 23} 573--610

\bibitem{Bongs2015}
Bongs K, Singh Y, Smith L, He W, Kock O, Świerad D, Hughes J, Schiller S,
  Alighanbari S, Origlia S, Vogt S, Sterr U, Lisdat C, Le~Targat R, Lodewyck J,
  Holleville D, Venon B, Bize S, Barwood G~P, Gill P, Hill I~R, Ovchinnikov
  Y~B, Poli N, Tino G~M, Stuhler J and Kaenders W 2015 {\em Comptes Rendus
  Physique\/} {\bf 16} 553--564

\bibitem{Moudrak2008}
Moudrak A, Klein H and Eissfeller B 2008 {\em InsideGNSS\/} {\bf Sept./Oct.}
  45--50

\bibitem{Delva2013}
Delva P and Lodewyck J 2013 {\em Acta Futura\/} {\bf 7} 67--78

\bibitem{Chiaverini2005}
Chiaverini J, Blakestad R~B, Britton J, Jost J~D, Langer C, Leibfried D, Ozeri
  R and Wineland D~J 2005 {\em Quant. Inf. and Comp.\/} {\bf 5} 419--439

\bibitem{Eltony2015}
Eltony A~M, Gangloff D, Shi M, Bylinskii A, Vuletić V and Chuang I~L 2015 {\em
  arXiv:1502.05739 [physics, physics:quant-ph]\/} ArXiv: 1502.05739

\bibitem{Turchette2000}
Turchette Q~A, Kielpinski, King B~E, Leibfried D, Meekhof D~M, Myatt C~J, Rowe
  M~A, Sackett C~A, Wood C~S, Itano W~M, Monroe C and Wineland D~J 2000 {\em
  Phys. Rev. A\/} {\bf 61} 063418

\bibitem{Dubessy2009}
Dubessy R, Coudreau T and Guidoni L 2009 {\em Phys. Rev. A\/} {\bf 80} 031402

\bibitem{Hite2012}
Hite D~A, Colombe Y, Wilson A~C, Brown K~R, Warring U, Jördens R, Jost J~D,
  McKay K~S, Pappas D~P, Leibfried D and Wineland D~J 2012 {\em Phys. Rev.
  Lett.\/} {\bf 109} 103001

\bibitem{McKay2014}
McKay K~S, Hite D~A, Colombe Y, Jördens R, Wilson A~C, Slichter D~H, Allcock
  D~T~C, Leibfried D, Wineland D~J and Pappas D~P 2014 {\em arXiv:1406.1778
  [physics, physics:quant-ph]\/} ArXiv: 1406.1778

\bibitem{Du2004}
Du S, Squires M~B, Imai Y, Czaia L, Saravanan R~A, Bright V, Reichel J,
  H\"ansch T~W and Anderson D~Z 2004 {\em Phys. Rev. A\/} {\bf 70} 053606

\bibitem{Farkas2010}
Farkas D~M, Hudek K~M, Salim E~A, Segal S~R, Squires M~B and Anderson D~Z 2010
  {\em App. Phys. Lett.\/} {\bf 96} 093102

\bibitem{House2008}
House M~G 2008 {\em Phys. Rev. A\/} {\bf 78} 033402

\bibitem{Wesenberg2008}
Wesenberg J~H 2008 {\em Phys. Rev. A\/} {\bf 78} 063410

\bibitem{Deslauriers2006}
Deslauriers L, Olmschenk S, Stick D, Hensinger W~K, Sterk J and Monroe C 2006
  {\em Phys. Rev. Lett.\/} {\bf 97} 103007

\bibitem{McLoughlin2011}
McLoughlin J~J, Nizamani A~H, Siverns J~D, Sterling R~C, Hughes M~D, Lekitsch
  B, Stein B, Weidt S and Hensinger W~K 2011 {\em Phys. Rev. A\/} {\bf 83}
  013406

\bibitem{Pyka2013}
Pyka K, Herschbach N, Keller J and Mehlst\"aubler T~E 2013 {\em App. Phys. B\/}
   1--11

\bibitem{Lee2014}
Lee M~W, Jarratt M~C, Marciniak C and Biercuk M~J 2014 {\em Opt. Expr.\/} {\bf
  22} 7210

\bibitem{Saleh2015}
Saleh K, Millo J, Didier A, Kersal\'e Y and Lacro\^ute C 2015 {\em Appl.
  Opt.\/} {\bf 54} 9446--9449

\bibitem{Didier2015}
Didier A, Millo J, Grop S, Dubois B, Bigler E, Rubiola E, Lacro\^ute C and
  Kersal\'e Y 2015 {\em Appl. Opt.\/} {\bf 54} 3682--3686

\bibitem{Rovera1994}
Rovera G~D, Santarelli G and Clairon A 1994 {\em Rev. of Sci. Instr.\/} {\bf
  65} 1502--1505

\end{thebibliography}

\end{document}